\documentclass[twocolumn,amsmath,amssymb,floatfix,10pt,aps,pre]{revtex4-1}

\newcommand{\bvec}[1]{{\mathbf{\string#1} }}
\newcommand{\upd}{\mathrm{d}}
\usepackage{graphicx}
\usepackage{amssymb,amsfonts,amsmath,mathrsfs}
\usepackage{color}
\usepackage{ulem}

\usepackage{relsize}

\usepackage{upgreek}

\newcommand{\matX}[4]{\ensuremath{\left(\begin{array}{c}#1\\#3\end{array}\begin{array}{c}#2\\#4\end{array}\right)}}

\newcommand{\rr}{\bvec{r}}
\newcommand{\nab}{\boldsymbol{\nabla}}

\newcommand{\taua}{\tau_\text{a}}

\newcommand{\Da}{\mathcal{D}_\text{a}}

\DeclareSymbolFont{matha}{OML}{txmi}{m}{it}
\DeclareMathSymbol{\varv}{\mathord}{matha}{118}
\newcommand{\Vext}{\varv}

\newcommand{\di}{\mathfrak{d}}

\begin{document}

\title{Effective equilibrium states in mixtures of active particles driven by colored noise}
\author{Ren\'e Wittmann}
\email{rene.wittmann@unifr.ch}
 \affiliation{Department of Physics, University of Fribourg, CH-1700 Fribourg, Switzerland}
\author{J.\ M.\ Brader}\affiliation{Department of Physics, University of Fribourg, CH-1700 Fribourg, Switzerland}
\author{A. Sharma}
\affiliation{Leibniz-Institut f\"ur Polymerforschung Dresden, D-01069 Dresden, Germany}
\author{U.\ Marini Bettolo Marconi}
\affiliation{Scuola di Scienze e Tecnologie, 
Universit\`a di Camerino, Via Madonna delle Carceri, I-62032, Camerino, INFN Perugia, Italy}
\date{\today}

\begin{abstract} 
 We consider the steady-state behavior of pairs of active particles having different persistence times and diffusivities.
To this purpose we employ the active Ornstein-Uhlenbeck model, where the 
particles are driven by colored noises with exponential correlation functions whose intensities and correlation times vary from species to  species.
 By extending Fox's theory to many components, we derive
by functional calculus an approximate Fokker-Planck equation for the  configurational distribution function of the system.
After illustrating the predicted distribution
in the solvable case of two particles interacting via a harmonic potential, we consider systems of particles repelling through inverse power-law potentials.
We compare the analytic predictions
to computer simulations for such soft-repulsive interactions in one dimension,
and show that at linear order in the persistence times the theory is satisfactory.
 This work provides the toolbox to qualitatively describe many-body phenomena, such as demixing and depletion, by means of effective pair potentials.
\end{abstract}

\maketitle

\section{Introduction}

The study of active particles has recently attracted rapidly increasing attention 
of scientists belonging to different disciplines due to the current interest in the physical 
principles governing the behavior of fish schools, herds of animals, bacteria, collections of cells 
and/or man-made active colloids~\cite{REVmarchetti2013,REVelgeti2015,REVbechinger2016}. 
All these systems in order to move convert energy via metabolic or chemical reactions and are thus out of equilibrium~\cite{ebbenshowse2010}.
On the theory side, many fundamental aspects of active systems can be described with some
minimalistic models based on spherical particles~\cite{REVromanczuk2012,REVmarchetti2016}.
Even in the absence of attractive interactions such particles can exhibit intriguing individual and collective behavior, induced solely by their persistent motion,
such as the accumulation at the system boundary \cite{Yang,ni2015}, 
the separation into a dilute and a dense phase \cite{buttinoni2013,CatesReview} and wetting or
capillary condensation transitions~\cite{ni2015,wittmannbrader2016}.

While the majority of studies are concerned with systems whose constituents are all identical,
in real situations it is common to observe assemblies of active particles of different nature.
The obvious question is how does the heterogeneity affect the collective behavior of such mixtures~\cite{REVbechinger2016}.
 For example, doping a passive fluid with a small number of active particles 
significantly alters its structural and dynamical
properties by supporting the formation of clusters \cite{kuemmel2015} 
and, at higher densities, crystallization \cite{NiCohenstuart2014,kuemmel2015}.
On the other hand, active dopants with a short persistence length were reported to aggregate in cages~\cite{tanaka2017dope}.
On immersing large 
colloids into a bath of smaller active colloids,
the former effectively are rendered active, which becomes manifest through activity-enhanced 
diffusivities~\cite{wu2000,valeriani2011} and 
depletion forces~\cite{angelani2011,harder2014,ni2015,smallenburg2015}.
Employing shape-anisotropic colloids~\cite{mallory2014,smallenburg2015,maggi2016motor}
 or manufacturing activity gradients~\cite{merlitz2017},
these effects can be used to generate directed motion of the colloids. 
Moreover, the segregation between two passive species has been reported as the result of coupling only one species to the active bath~\cite{DasPolley2016segregation}.

The mixtures described so far consist of
species which differ in their shape (interaction potential) and particle number.
Seeking for the closest analogy to the motility-induced phenomena observed for a single active species,
we are particularly interested in particles solely distinguished by a difference in their activity.
The most intriguing feature of such multicomponent active systems is their capability to demix, 
which cannot be attributed to the physical mechanisms also present in equilibrium mixtures.
Some recent investigations have focused on such a binary mixture of an active
and a passive species~\cite{wysocki2016propagatingAP,stenhammar2015activity,vandermeer2017} 
or particles with different finite activities~\cite{takatori2015theory,kumari2017demixing,smrek2017,vandermeer2017}.
Quite intuitively, active phase separation phenomena can be described using the 
concept of an effective temperature, enhanced by activity~\cite{palacci2010}, 
which has also been applied to mixtures~\cite{han2017effectivetemp,grosberg2015nonequilibrium,takatori2015theory}.
Relatedly, the demixing of particles with the same mobility but different diffusion coefficients has been recently reported \cite{weber2016}.

The model of active particles propelled by so-called Ornstein-Uhlenbeck processes (OUPs) 
provides a convenient starting point of many theoretical studies \cite{fodor2016,szamel2014,szamel2015,fily2012,flenner2016nonequilibrium,szamel2017evaluating},
 since their equations of motion do not resolve the orientational degrees of freedom.
A minimalistic strategy is based on the
multidimensional generalizations of the unified colored noise approximation (UCNA)~\cite{ucna1,ucna2} or a similar approach by Fox~\cite{fox1,fox2},
see Ref.~\cite{activePair} for a detailed comparison.
This procedure yields an approximate Smoluchowski equation, which, in the steady state, admits an
analytic solution for the configurational probability distribution \cite{maggi2015sr},
and closed formulas for active pressure and interfacial tension \cite{marconi2016,activePressure,marconiExactpressure2017}.
The former allows to define effective interaction potentials,
which can be directly used to determine density profiles \cite{marconi2015,activePair,filybaskaranEETcurved2017}
and rate equations \cite{sharma2016} of individual ideal particles
and, when implemented in equilibrium liquid-state theory, 
the structure and phase behavior of interacting systems \cite{marconi2016mp,faragebrader2015,wittmannbrader2016}.
The described effective equilibrium approach is most accurate in one spatial dimension and for small persistence time,
which can be explicitly verified by studying exactly solvable models \cite{marconiExactpressure2017}.

In the present work we take the effective equilibrium model to the next level 
by further generalizing the Fox approach, which turns out to be more promising, to mixtures of different active particles. 
 For two particles, we demonstrate that the theory yields a pairwise potential which agrees well with simulations of active OUPs.
The paper is organized as follows.
 The description of the model and the generalization of the Fox approach are presented in Sec.~\ref{Theory}. 
We then verify in Sec.~\ref{Test} the accuracy of the theory at linear order in the persistence times
by studying an exactly solvable harmonic problem.
In Sec.~\ref{Probabilityd} we discuss the configurational probability
distribution and compare to numerical results. 
Finally, in Sec.~\ref{Conclusions}, we draw some conclusions regarding the meaning of our results on the many-particle level.

\section{Theory}
\label{Theory}

In order to study a system of $N$ active particles in $\di$ spatial dimensions having species dependent diffusivities $D_\text{a}^{(i)}$ and persistence times $\taua^{(i)}$
where  the index $i\in\{1\ldots N\}$,
we generalize the  microscopic one-component active OUPs 
model of Refs.~\cite{maggi2015sr,marconi2015,activePair} to the case where 
different types of Gaussian stochastic driving terms are present. To do so, we introduce
a component-wise notation (compare, e.g., Ref.~\cite{marconi2015}) for $\di N$-dimensional arrays $x_\upalpha(t)$ 
denoting the coordinates of the particles evolving according to
\begin{align}
\dot{x}_\upalpha(t)=D_\text{t}\beta F_\upalpha(x_1,x_2,\ldots,x_{\di N})+\chi_\upalpha(t) \,,
\label{eq_FOXapp1}
\end{align}
where $\upalpha\in\{1\ldots \di N\}$, $F_\upalpha$ is a conservative force due to passive interactions and
$D_\text{t}\beta$ is the inverse friction coefficient (related to the translational diffusivity
 $D_\text{t}$ in a Brownian system).

The Gaussian stochastic noise $\chi_\upalpha(t)$ evolves in time according to
\begin{align}
\dot \chi_\upalpha(t) =- \frac{1}{\taua^{(\upalpha)}} \chi_\upalpha(t) + \frac{ \sqrt{\Da^{(\upalpha)}}}{ \taua^{(\upalpha)}} \,\xi_\upalpha(t)  
\label{eq_OUPsdef}
\end{align}
with the white noise $\xi_\upalpha(t)$, which has the time correlator $\langle\xi_\upalpha(t)\xi_\upbeta(t')\rangle\!=\!2D_\text{t}\delta_{\upalpha\upbeta}\delta(t-t')$, and $\Da^{(\upalpha)}=D_\text{a}^{(\upalpha)}/D_\text{t}$.
It has zero average and the tensorial time correlator 
\begin{equation}
C_{\upalpha\upbeta}(t-t'):=\langle\chi_\upalpha(t)\chi_\upbeta(t')\rangle
 \!=\!\frac{D_\text{a}^{(\upalpha)}}{\taua^{(\upalpha)}}\delta_{\upalpha\upbeta}e^{-\frac{|t\!-\!t'|}{\taua^{(\upalpha)}}}\,.
 \label{eq_OUPsCORRapp}
\end{equation}
The probability distribution functional of $\chi_\upalpha(t)$ has the Gaussian representation:
\begin{align}
P_N[\{\chi_\upalpha\}]\!\propto\!\exp\!\!\left(\!-\frac{1}{2}\iint\!\upd s\,\upd s'\sum_{\upalpha\upbeta}
\chi_\upalpha(s)K_{\upalpha\upbeta}(s-s')\chi_\upbeta(s')\!\right)
\label{eq_FOXapp2}
\end{align}
 and is equipped with a tensorial kernel $K_{\upalpha\upbeta}(t-t')$,
the inverse of $C_{\upalpha\upbeta}$.

\subsection{Fokker-Planck equation}

In appendix~\ref{appendix:foxapproach}, by extending Fox's approximation to an arbitrary number of (active) components, we show that the configurational distribution $f_N(\{x_\upalpha\},t)$ of positions $x_\upalpha$ of particles
evolves according to the following Fokker-Planck equation:
\begin{align}
&\frac{\partial f_N(\{x_\upalpha\},t)}{\partial t}=
-\sum_\upbeta\frac{\partial}{\partial x_\upbeta} \Bigl(
D_\text{t}\beta F_{\upbeta}(\{x_\upalpha\})f_N(\{x_\upalpha\},t)\nonumber\\ &\ \ \ \ \ \ \ \ \ \ \ 
-D_\text{a}^{(\upbeta)} \sum_\upgamma  \frac{\partial}{\partial x_\upgamma} f_N( \{x_\upalpha\},t) 
\Gamma _{\upgamma\upbeta}^{-1}( \{x_\upalpha\}) \Bigr) \!\!\!\!\!\!\!\!\!
\label{eq_FOXappFP11}
\end{align}
with friction matrix
\begin{align}
\Gamma_{\upgamma \upbeta}= \delta_{\upgamma \upbeta}-\taua^{(\upbeta)} 
D_\text{t}  \beta   \partial_\upbeta F_\upgamma
\label{eq_gammaF}
\end{align}
 and the short notation $\partial_\upbeta\!:=\!\partial/\partial x_\upbeta$ for the partial derivative employed here and in the following.

Intriguingly, the generalized Unified Colored Noise Approximation (UCNA)~\cite{ucna1,ucna2}
gives rise to a friction matrix
\begin{align}
\Gamma^{\text{ucna}}_{\upgamma \upbeta}= \delta_{\upgamma \upbeta}-
\taua^{(\upgamma)} D_\text{t}  \beta \partial_\upbeta F_\upgamma
=\Gamma_{\upbeta\upgamma}\,,
\label{eq_gammaU}
\end{align}
which is the transpose of the Fox result in Eq.~\eqref{eq_gammaF}, since, for a conservative force, we have $\partial_\upbeta F_\upgamma=\partial_\upgamma F_\upbeta$.
 In either case, $\Gamma_{\upbeta\upgamma}=\Gamma_{\upgamma\upbeta}$ only holds if the particles labeled $\upbeta$ and $\upgamma$ belong to the same species.
 Most importantly, we find that 
 $
 \Da^{(\upbeta)}\,(\Gamma^{\text{ucna}})_{\upgamma\upbeta}^{-1}\!=\!\Da^{(\upbeta)}\,\Gamma^{-1}_{\upbeta\upgamma}
 $
does in general not even correspond to the transpose of
the Fox expression $\Da^{(\upbeta)}\,\Gamma^{-1}_{\upgamma\upbeta}$ entering in Eq.~\eqref{eq_FOXappFP11}, contrasting the relation in Eq.~\eqref{eq_gammaU}. 
Therefore, UCNA and Fox only share the same steady state in a (non-thermal) one-component system.
As detailed later, differences between these two approaches arise even at linear order in the persistence times.

While the steady-state condition $\partial f_N/\partial t=0$ requires in general the vanishing of the divergence of the probability current, the 
condition of detailed balance involves the vanishing of all components of the probability current, so that 
from Eq.~\eqref{eq_FOXappFP11} we find
\begin{align}
f_N \Bigl(\beta F_{\upbeta}  -   \Da^{(\upbeta)}\sum_\upgamma \partial_\upgamma  \Gamma_{\upgamma\upbeta}^{-1} \Bigr)
 =\Da^{(\upbeta)} \sum_\upgamma   \Gamma _{\upgamma\upbeta}^{-1} \partial_\upgamma f_N \, .
\label{eq_FOXdetailed2}
\end{align}
In the case of mixtures considered here, the solution of such an equation is not known, but it reduces to  a Boltzmann-like distribution
for a single species~\cite{maggi2015sr}.
 Only if all diffusivities $\Da^{(\upbeta)}$ are equal, Eq.~\eqref{eq_FOXdetailed2} is fulfilled by $\Gamma^{\text{ucna}}_{\upgamma \upbeta}$
but not by $\Gamma_{\upgamma \upbeta}$, which shows that Fox theory is, in general, less equilibrium-like than UCNA.

Although the derivation of Eq.~\eqref{eq_FOXappFP11} is valid for any dimensionality,
 we will restrict ourselves to $\di\!=\!1$ spatial dimensions in the remainder of this work,
 so that the Greek indices become particle labels.
Some more general results for higher dimensions are stated in appendix~\ref{appendix:cartesian}.
Assuming that the conservative force arises from an interaction potential $u_{\upgamma\upalpha}(x_\upgamma-x_\upalpha)$ between each pair of particles,
we can recast Eq.~\eqref{eq_gammaF} as
\begin{align}
\Gamma_{\upgamma \upbeta}=\delta_{\upgamma \upbeta}+\tau^{(\upbeta)} \partial_\upbeta\partial_\upgamma
\sum_{\upalpha\neq \upgamma}^N \beta u_{\upgamma\upalpha}\,,
\label{eq_gamma0}
\end{align}
where we define the dimensionless persistence times~\cite{activePair}
\begin{align}
\tau^{(\upalpha)}= D_\text{t} \taua^{(\upalpha)}/d^2=D_\text{t} \taua^{(\upalpha)}\,,
\label{eq_tauj}
\end{align}
 setting the unit of length $d$, entering in the specification of the pair potential, to unity.
 Since this friction matrix enters Eq.~\eqref{eq_FOXappFP11} in a non-trivial way, our further strategy involves approximating many-body by pairwise quantities, 
which we explore in the following by further restricting ourselves to $N=2$ particles, i.e., $\upalpha\in\{1,2\}$.

\section{An elementary test}
\label{Test}

 As a first step, we consider perhaps the simplest
model of interacting particles which lends itself to an analytic solution and 
may serve as a benchmark for our theory. 
An elastic dimer in a one-dimensional well and 
subject to two different colored Gaussian baths
is represented by two particles 
 mutually coupled by a harmonic potential $\beta u(x_1-x_2)=\alpha^2 (x_1-x_2)^2/2$
and each confined by a harmonic external potential $\beta\Vext(x_\upalpha)=\omega^2 x^2_\upalpha/2$.
Their dynamics is described by the  evolution equations
\begin{align}
\dot{x}_\upalpha(t)=D_\text{t}\beta F_\upalpha(x_1,x_2)+\chi_\upalpha(t)\,,
\label{eq_FOXosc1}
\end{align}
where  $F_\upalpha\!=\!- \partial_\upalpha(\Vext (x_\upalpha)+u(x_\upalpha -x_\upbeta))$.
The friction $1/( D_\text{t}\beta)$ is the same for both species, but the parameters $\taua^{(\upalpha)}$ and $\Da^{(\upalpha)}$ 
characterizing the stochastic driving term $\chi_\upalpha(t)$, defined in Eq.~\eqref{eq_OUPsdef},
are different.
Explicitly we have
\begin{align}
 &&\dot x_1=- D_\text{t}(\omega^2 x_1+\alpha^2 (x_1-x_2))+\chi_1(t)\,,
 \label{x1dot} \\
 &&\dot x_2=- D_\text{t}(\omega^2 x_2-\alpha^2 (x_1-x_2))+\chi_2(t)\,.
\label{x2dot}
\end{align}
This model has been first used in statistical mechanics
in the context of the virial theorem by Riddell and Uhlenbeck \cite{RiddellUhlenbeck}
 and recently by one of us \cite{marconiExactpressure2017} in the framework of active systems.

The model, being linear, can be solved analytically  by direct integration of equations of motion.
It is convenient to switch to collective variables $q\!=\!x_1-x_2$ and $Q\!=\!(x_1+x_2)/2$ and to the renormalized spring constant $\Omega^2\!=\!\omega^2+2\alpha^2$.
The steady-state equal-time pair correlations read
\begin{align}
\langle Q(t) Q(t)  \rangle&=
\frac{1}{4 \omega^2} 
\Bigl( 
\frac{\Da^{(1)}}  {1+\tau^{(1)}\omega^2}+ 
\frac{\Da^{(2)}}   {1+\tau^{(2)} \omega^2} \Bigr )\,,
\cr
\langle q(t) q(t)  \rangle&=
\frac{1 }{  \Omega^2} \Bigl( 
\frac{\Da^{(1)}}   {1+\tau^{(1)} \Omega^2}+ 
\frac{\Da^{(2)}}   {1+\tau^{(2)} \Omega^2} \Bigr )\,.
\label{eq_qqE} 
\end{align}
Such an exact result will now be used to compare with the generalized Fox approximation.

\subsection{Fox's approximation for two oscillators \label{sec_dimerFOX}}

We now compute the averages featuring in Eq.~\eqref{eq_qqE} using the approximate Fox theory,
 as described in appendix~\ref{appendix:foxapproach}, for two components.
For the case of the Riddell-Uhlenbeck model,
let us introduce  the following symbols to shorten the notation:
$D_\text{t}\beta F_\upalpha =\sum_{\upbeta} M_{\upalpha\upbeta} x_\upbeta$ with
 $M_{11}\!=\!M_{22}\!=\!-D_\text{t}(\omega^2+\alpha^2)$ and
$M_{12}\!=\!M_{21}\!=\! D_\text{t}\alpha^2$. We write the evolution equations under the form:
\begin{align}
\frac{\partial }{\partial t}\langle x_\upalpha x_\upbeta \rangle= \sum_\upgamma [M_{\upalpha \upgamma} \langle x_\upgamma x_\upbeta\rangle + M_{\upbeta\upgamma} \langle x_\upgamma x_\upalpha\rangle]
&\cr+ D_\text{t} [
\Da^{(\upbeta)} \Gamma^{-1}_{\upalpha \upbeta}
+ \Da^{(\upalpha)}\Gamma^{-1}_{\upbeta\upalpha}]
\label{eq_average}
\end{align}
with
 $\Gamma_{\upalpha\upbeta}\!=\!\delta_{\upalpha\upbeta}-\tau^{(\upbeta)} \beta 
\partial_\upbeta F_\upalpha$ according to Eq.~\eqref{eq_gammaF}.
 Since the  motion is confined by the external potential, the chosen observables  are limited, and  by the ergodic theorem~\cite{falasco2015mesoscopic}
the left-hand side of Eq.~\eqref{eq_average} vanishes as $t\!\to\!\infty$ and the system approaches the steady state.
In this case, we obtain a simple set of linear equations for the correlators $\langle x_\upalpha x_\upbeta\rangle$, which
 can be easily solved and we find the general expressions
 \begin{align}
 \langle Q(t) Q(t) 
\rangle&=
\frac{1}{4  \omega^2} 
\sum_{\upalpha\upbeta} \Da^{(\upbeta)}\Gamma^{-1}_{\upalpha\upbeta}\,,\cr
 \langle q(t) q(t) 
\rangle&=\frac{1 }{\Omega^2} \sum_{\upalpha\upbeta} (-1)^{\upalpha+\upbeta}\, \Da^{(\upbeta)}\Gamma^{-1}_{\upalpha\upbeta}
\label{eq_qqFgen} 
\end{align}
 for the equal-time correlations of the collective variables.
 
Explicitly, the inverse matrix $\Gamma_{\upalpha\upbeta}^{-1}$ reads
\begin{align}
&\Gamma^{-1}_{11}=|\Gamma|^{-1}\left(1+\tau^{(2)} (\omega^2+\alpha^2)\right)\rightarrow1-\tau^{(1)} (\omega^2+\alpha^2)\,, \nonumber\\
&\Gamma^{-1}_{22}=|\Gamma|^{-1}\left(1+ \tau^{(1)}  (\omega^2+\alpha^2)\right)\rightarrow1- \tau^{(2)}  (\omega^2+\alpha^2)\,, \nonumber\\
 &\Gamma^{-1}_{12}=|\Gamma|^{-1}\,\tau^{(1)}  \alpha^2\rightarrow\tau^{(2)}  \alpha^2\,, \nonumber\\
& \Gamma^{-1}_{21}=|\Gamma|^{-1}\,\tau^{(2)}  \alpha^2\rightarrow\tau^{(1)}  \alpha^2\,, \! \! \! \! \! \! \label{eq_GAMMAlinH}
\end{align}
where $|\Gamma|=1+(\tau^{(1)}+\tau^{(2)})(\omega^2+\alpha^2)+\tau^{(1)}\tau^{(2)}(\omega^4+2\omega^2\alpha^2)$ is the determinant of $\Gamma_{\upalpha\upbeta}$
and the last expressions are valid to first order in $\tau^{(\upalpha)}$.
Plugging these expressions into Eq.~\eqref{eq_qqFgen} and comparing with the corresponding exact correlators $\langle QQ\rangle$ and $\langle qq\rangle$,
we recognize that these cannot be expressed in terms of a single spring constant, $\omega^2$ or $\Omega^2$, as in Eq.~\eqref{eq_qqE}.
However, we can identify the common leading terms
\begin{align}\!\!\!\!
 \langle QQ 
\rangle&\!=\!
\frac{1}{4  \omega^2} 
\Bigl( 
\Da^{(1)}  (1-\tau^{(1)}\omega^2)+ 
\Da^{(2)}   (1-\tau^{(2)} \omega^2)\! \Bigr )\,,\!\!\!\cr \!\!\!\!
 \langle qq 
\rangle&\!=\!
\frac{1 }{\Omega^2} \Bigl( 
\Da^{(1)}  (1-\tau^{(1)} \Omega^2)+ 
\Da^{(2)}   (1-\tau^{(2)} \Omega^2)\! \Bigr )\!\!\!
\label{eq_qqF} 
\end{align}
of both results,
which means that Fox's theory is exact to linear order in $\tau^{(\upalpha)}$.

One can easily see from comparing Eq.~\eqref{eq_gammaF} and Eq.~\eqref{eq_gammaU} that the UCNA results for the off-diagonal friction matrix elements in Eq.~\eqref{eq_GAMMAlinH} 
and for the correlators in Eq.~\eqref{eq_qqFgen} are different from those found in 
the present treatment, based on the Fox approximation, even in the small $\tau^{(\upalpha)}$ limit, given by Eq.~\eqref{eq_qqF}.
 This observation makes sense regarding the nature of the derivation in appendix~\ref{appendix:foxapproach},
whereas the UCNA becomes uncontrolled when more than one time scale is involved.
Therefore, we shall not digress to further discuss the multicomponent UCNA equations.

\section{Probability distribution function}
\label{Probabilityd}

Having seen that a system of two active particles with different diffusivities and (small) 
persistence times are accurately described by the multicomponent Fox approach,
our goal is to describe a system with $\kappa$ components in $\di$ dimensions.
As shown in Sec.~\ref{Theory}, the effective Fokker-Planck equation~\eqref{eq_FOXappFP11} does, in general, not admit current-free steady states.
 It is thus not possible to identify 
a Boltzmann-like expression for the configurational probability distribution $f_N(\bvec{r}^N)$,
which fulfills Eq.~\eqref{eq_FOXdetailed2}.
However, although such a solution is known analytically for a one-component system~\cite{maggi2015sr},
further approximations are unavoidable in order to turn this advantage into a workable theory for the many-body system~\cite{activePair}.

To make progress, we aim to provide the recipe to reconstruct a pairwise-additive approximation
for the many-body effective interaction potential $\beta \mathcal{H}_{[N]}\!=\!-\ln f_N$, associated with a
(presumably) Boltzmann-like distribution $f_N$~\cite{activePair,faragebrader2015}.
Given this objective, we consider only $N\!=\!2$ particles in the first place
and determine the solution for $f_2$.
As before, we only discuss a one-dimensional system and show some more general formulas in appendix~\ref{appendix:cartesian}.

For completeness, we also consider the case where some particles are subject to an additional translational Brownian white noise 
(referred to as thermal noise in the following) entering in Eq.~\eqref{eq_FOXapp1},
 which facilitates establishing the connection to mixtures of active Brownian particles.
 Within the Fox approximation, this simply amounts to adding the term 
$I_\text{t}^{(\upbeta)} D_\text{t} \partial^2_\upbeta f_N$
 on the right-hand-side of Eq.~\eqref{eq_FOXappFP11} \cite{faragebrader2015,sharma2016,activePair},
 where $I_\text{t}^{(\upbeta)}$ takes the values 1 in the presence and 0 in the absence of thermal noise acting on particle $\upbeta\in\{1,2\}$.

\subsection{Two-particle current \label{sec_twopartcurrent}}

Explicitly, for $N\!=\!2$, we can rewrite Eq.~\eqref{eq_FOXappFP11} as $\partial f_2/\partial t=-D_\text{t}\sum_{\upbeta}\partial_{\upbeta}J_\upbeta$.
The two-body probability current reads
\begin{align}
J_{\upbeta}(x)=\beta F_\upbeta(x) f_2(x)-\sum_{\upgamma}\partial_{\upgamma}(\mathcal{D}_{\upgamma\upbeta}(x)f_2(x))\,,
\label{eq_ss1}
\end{align}
where all quantities only depend on the (relative) distance $x\!=\!x_1-x_2$ and $\mathcal{D}_{\upgamma\upbeta}=\delta_{\upgamma\upbeta}I_\text{t}^{({\upbeta})}+
\Da^{({\upbeta})}\,\Gamma_{\upgamma\upbeta}^{-1}$ is the effective $2\times2$ diffusion tensor, 
which can be written in matrix notation, introducing the common derivative operator $\partial_x\equiv\partial_1\!=\!-\partial_2$, as
\begin{align}
\mathcal{D}(x)=\matX{
\frac{I_\text{t}^{(1)}+\Da^{(1)}\left(1+\tau^{(2)}\partial_x^2 \beta u(x)\right)}{1+(\tau^{(1)}+\tau^{(2)})\partial_x^2 \beta u(x)}}{
\frac{\Da^{(2)}\tau^{(2)}\partial_x^2 \beta u(x)}{1+(\tau^{(1)}+\tau^{(2)})\partial_x^2 \beta u(x)}}{
\frac{\Da^{(1)}\tau^{(1)}\partial_x^2 \beta u(x)}{1+(\tau^{(1)}+\tau^{(2)})\partial_x^2 \beta u(x)}}{
\frac{I_\text{t}^{(2)}+\Da^{(2)}\left(1+\tau^{(1)}\partial_x^2 \beta u(x)\right)}{1+(\tau^{(1)}+\tau^{(2)})\partial_x^2 \beta u(x)}},
\label{eq_Deff2}
\end{align}
 where $\upgamma$ denotes the column and $\upbeta$ denotes the row.

The steady-state condition $\partial f_2/\partial t=0$ is equivalent to
\begin{equation}
-\partial_xJ_1+\partial_xJ_2=0\,. 
\label{eq_ss1ld}
\end{equation}
Explicitly, the two currents are given by:
\begin{align}
{J}_1(x)=&f_2(x) \,\partial_x(-\beta u(x)-\mathcal{D}_{11}+\mathcal{D}_{21})\cr&+(-\mathcal{D}_{11}+\mathcal{D}_{21})\,\partial_x f_2\,,\cr
{J}_2(x)=& f_2(x)\,\partial_x (\beta u(x)-\mathcal{D}_{12}+\mathcal{D}_{22}) \cr&+(-\mathcal{D}_{12}+\mathcal{D}_{22})\,\partial_x f_2\,.
\label{eq_j2} 
\end{align}
Considering two members of the same species,
where $\mathcal{D}_{11}\!=\!\mathcal{D}_{22}$ and $\mathcal{D}_{21}\!=\!\mathcal{D}_{12}$,
we easily see that Eq.~\eqref{eq_ss1ld} is trivially fulfilled.
In general, this zero-divergence condition is satisfied by  
$J_1=J_2+X$, where $X$ is a constant.

 Suppose the two particles belonging to different species interact with the same finite-range and symmetric pair potential $u(x)$ with $\lim_{x\to\pm\infty} u(x)=0$.
By subtracting the second current in Eq.~\eqref{eq_j2} from the first one we obtain
\begin{align}
X= -2f_2(x)\, \partial_x \Bigl(\beta u(x)+\mathcal{D}_\text{m}\Bigr)-2\mathcal{D}_\text{m}\partial_x f_2(x)\,,
\label{eq_bigX} 
\end{align}
 where  we defined
\begin{align}
\mathcal{D}_\text{m}(x):=\frac{1}{2}\big(\mathcal{D}_{11}-\mathcal{D}_{21}+\mathcal{D}_{22}-\mathcal{D}_{12}\big)\,.
\label{eq_D4} 
\end{align}
Such an inhomogeneous first order linear differential equation can be solved by introducing the so-called integrating factor
\begin{align}
\psi_2(x)=\exp \left( -\int_{-\infty}^x  \frac{\upd s}{\mathcal{D}_\text{m}(s)} \, \partial_s (\beta u(s)  +\mathcal{D}_\text{m}(s)) \right).
\label{eq_psi2}
\end{align}

Due to the symmetry of the pair potential
the integrating factor has the property $\lim_{x\to\pm\infty} \psi_2(x)=1$.
The general solution  of Eq.~\eqref{eq_bigX}  must be of the form
\begin{align}
f_2(x)=  \psi_2(x)\left( A -\frac{X}{2} \int_{-\infty}^x \frac{\upd s}{ \mathcal{D}_\text{m}(s)\,  \psi_2(s)}  \right),
\label{eq_f2} 
\end{align}
so that  $ \lim_{x\to-\infty} f_2(x)=A$.
Now, the value of the distribution function at infinity must be identical to the value at minus infinity
( $\lim_{x\to\infty }f_2(x)= A$ ) and we may conclude that
\begin{align}
A=\lim_{x\to\infty} \psi_2(x)\left( A -\frac{X}{2} \int_{-\infty}^x \frac{\upd s}{ \mathcal{D}_\text{m}(s)\, \psi_2(s)} \right).
\end{align}
The only solution is $X\!=\!0$, thus the two currents $J_1=J_2$ must be equal
and $f_2(x)\!=\!\psi_2(x)$ is given by Eq.~\eqref{eq_psi2}, i.e., $A\!=\!1$.
Combining Eq.~\eqref{eq_bigX} for $X\!=\!0$ with Eq.~\eqref{eq_j2} we find
\begin{align}
 J_1=J_2=\frac{f_2(x)}{2\mathcal{D}_\text{m}}&\left((\mathcal{D}_{11}-\mathcal{D}_{21}-(\mathcal{D}_{22} -\mathcal{D}_{12}))\,\partial_x\beta u(x)\right.\cr
&\ \ +(\mathcal{D}_{11}-\mathcal{D}_{21})\,\partial_x(\mathcal{D}_{22} -\mathcal{D}_{12})\nonumber\\
&\ \ \left.-(\mathcal{D}_{22} -\mathcal{D}_{12})\,\partial_x(\mathcal{D}_{11}-\mathcal{D}_{21})
\right).\!\!\!\!\!\!\!\!\!\!\!\!\!\!
\label{eq_J1J2}
\end{align}
One can see that $J_1\!=\!J_2\!=\!0$ only if $\mathcal{D}_{11}-\mathcal{D}_{21}\!=\!\mathcal{D}_{22} -\mathcal{D}_{12}\!=\!\mathcal{D}_\text{m}$, a condition which is realized when the 
two particles are  identical (or if we set $\Da^{(1)}\!=\!\Da^{(2)}$ and use the UCNA expression for $\mathcal{D}(x)$,
obtained by exchanging in the diagonal terms of Eq.~\eqref{eq_Deff2} $\tau^{(1)}$ with $\tau^{(2)}$).

\subsection{An example of non-zero partial currents}

To illustrate that the equality $J_1\!=\!J_2$ of the (non-vanishing) partial currents
established in Sec.~\ref{sec_twopartcurrent} 
 does not hold only for particles interacting via soft-repulsive potentials but also for other interactions, 
we consider two different active OUPs (without thermal noise)
bound by the harmonic pair potential $u(x)\!=\!\alpha^2 x^2/2$, cf., Sec.~\ref{Test}.
In this case, we must have $\lim_{x\to\pm\infty} \psi_2(x)\!=\!0$ for the integrating factor in Eq.~\eqref{eq_psi2},
because the potential is confining.
Therefore, the argumentation leading to Eq.~\eqref{eq_J1J2} is no longer justified.
However, we also know the form of the equal-time pair correlation $\langle  q(t) \,q(t)\rangle\!=\!\mathcal{D}_\text{m}/(\alpha^2)$ in the Fox approximation,
compare Eq.~\eqref{eq_qqFgen} with $\Omega^2\equiv2\alpha^2$ to Eq.~\eqref{eq_D4}.
We can thus write the steady-state probability distribution 
\begin{align}
  f_2(x)\sim \exp \Bigl(\frac{-x^2}{2 \,\langle  q \,q\rangle} \Bigr) \sim \exp \Bigl(\frac{-\alpha^2 }{2 \mathcal{D}_\text{m}} x^2\Bigr)
   \sim \exp \Bigl(\frac{-\beta u(x) }{ \mathcal{D}_\text{m}} \Bigr)
   \label{eq_f2explicit}
\end{align}
explicitly as a Gaussian.

It is easily verified that a distribution of the form of Eq.~\eqref{eq_f2explicit} gives $X\!=\!0$ in Eq.~\eqref{eq_bigX}, 
since $\mathcal{D}_\text{m}$ does not depend on $x$ for the employed potential.
We immediately get from Eq.~\eqref{eq_J1J2} the position-dependent currents
 \begin{align}
\!\!\! J_1&=J_2=\frac{f_2(x)}{2\mathcal{D}_\text{m}} (\mathcal{D}_{11}-\mathcal{D}_{21}-\mathcal{D}_{22} +\mathcal{D}_{12})\,\partial_x\beta  u(x)\cr
&=\alpha^2xf_2(x)\mbox{$\frac{\big((\tau^{(1)}-\tau^{(2)})(\Da^{(1)}+\Da^{(2)})\big)\alpha^2-\Da^{(1)}+\Da^{(2)}}{\big((\tau^{(1)}-\tau^{(2)})(\Da^{(1)}-\Da^{(2)})\big)\alpha^2-\Da^{(1)}-\Da^{(2)}}$}\,.
\label{eq_J1J2bis}
\end{align}  
    From these formulas it is easy to verify that the currents vanish for two identical particles with both
   $\tau^{(1)}\!=\!\tau^{(2)}$ and $\Da^{(1)}\!=\!\Da^{(2)}$.
   The full formula for the probability distribution $f_2(x)$ follows from the consideration in the following section,
   which hold for an arbitrary bare interaction potential.

\subsection{Effective potentials \label{sec_ueffexplicit}}

With the knowledge of the two-particle probability distribution $f_2(x)\!\equiv\!\psi_2(x)$, as given by Eq.~\eqref{eq_psi2},
one can define an effective force $F_{\upalpha}^\text{eff}$ on particle $\upalpha\in\{1,2\}$ according to $\beta F_{\upalpha}^\text{eff} f_2-\partial_{\upalpha}f_2\!=\!0$,
 which has the form of a steady-state condition in a passive system.
Note that defining an effective force as $\beta\tilde{F}_{\upalpha}^\text{eff}\!=\!\mathcal{D}_{1\upalpha}^{-1}J_1+\mathcal{D}_{2\upalpha}^{-1}J_2+\partial_{\upalpha}f_2$, 
which in the one-component system is equivalent to the first definition~\cite{activePair},
does not yield the same result as $\beta F_{\upalpha}^\text{eff}$ in the general case of a mixture with non-vanishing probability currents.
We thus derive by equating the currents in Eq.~\eqref{eq_j2} the effective pair interaction potential 
\begin{align}
\partial_x\beta u_{\mu\nu}^\text{eff}(x)=-\partial_x\ln f_2=\mathcal{D}_\text{m}^{-1}\partial_x\,(\beta u(x)+\mathcal{D}_\text{m})
\label{eq_ueffij} 
\end{align} 
between two members of species $\mu$ and $\nu$,
with $\mathcal{D}_\text{m}$ defined in Eq.~\eqref{eq_D4}. 
 In general, we can then represent the effective many-body interaction by adding up the pair potentials $u_{\mu\nu}^\text{eff}(x_{\upalpha\upbeta})$ for all components and corresponding particle positions.
For the purpose of demonstration, we consider $\mu=1$ and $\nu\in\{1,2\}$ in the following.

The effective potentials $u_{11}^\text{eff}$ between members of the same species
can be obtained in various ways, e.g., simply by requiring that either current vanishes in Eq.~\eqref{eq_j2}.
The explicit form of the $u_{11}^\text{eff}$ has been discussed in detail in Ref.~\cite{activePair}.
Without thermal noise, i.e., setting in Eq.~\eqref{eq_Deff2} $I_\text{t}^{(1)}\!=\!I_\text{t}^{(2)}\!=\!0$,
the closed analytical expression 
\begin{align}
\beta u^\text{eff}_{11}(x)&=\frac{\upbeta u(x)+\tau^{(1)}\left(\partial_x \beta u(x)\right)^2}{\Da^{(1)}}-\ln\!\left|E^{(1)}_2(x)\right| 
\label{eq_ueffUCNA0}
\end{align}
 can also be found from integrating Eq.~\eqref{eq_ueffij}, where
\begin{align}
 E^{(\mu)}_2(x)=1+2\tau^{(\mu)} \partial^2_x\beta u(x)
 \label{eq_EV}
\end{align}
  denote the Eigenvalues of $\Gamma_{\upalpha\upbeta}$ from Eq.~\eqref{eq_gamma0}, evaluated for two particles of the same species $\mu$.
 In this special case, $E^{(\mu)}_2$ is equivalent (up to the factor $\Da^{(\mu)}$) to the 
 Eigenvalues of the more general (inverse) diffusion tensor $\mathcal{D}^{-1}_{\upalpha\upbeta}$ from Eq.~\eqref{eq_Deff2}.

 In the most general case of two different species with thermal noise present, 
 we can express $u_{12}^\text{eff}$ from Eq.~\eqref{eq_ueffij}
 with help of the Eigenvalues 
 \begin{align}
\mathcal{E}_2(x)=\frac{\bar{E}_2(x)}{\bar{\mathcal{D}}_\text{a}-\Delta\Da\,\Delta E_2(x)+\bar{I}_\text{t}\bar{E}_2(x)}
 \label{eq_EVgen}
\end{align}
of $\mathcal{D}_\text{m}^{-1}$ from Eq.~\eqref{eq_D4} with Eq.~\eqref{eq_Deff2}
and the average and deviatoric parameters
 \begin{align}
 \bar{\Xi}:=\frac{\Xi^{(1)}+\Xi^{(2)}}{2}\,,\ \ \  \Delta \Xi:=\frac{\Xi^{(1)}-\Xi^{(2)}}{2}\,,
 \label{eq_avpars}
\end{align}
where $\Xi^{(\mu)}$ represents the Eigenvalues $E^{(\mu)}_2$ from Eq.~\eqref{eq_EV}, the persistence times $\tau^{(\mu)}$,
the active diffusivities $\Da^{(\mu)}$ or the characteristic functions $I_\text{t}^{(\mu)}$ of thermal noise.
 It is a known problem of the present theory that, already for the one-component system,
 the effective potential is only well defined if the Eigenvalues of the diffusion tensor are strictly positive.
 Systems for which this validity criterion is fulfilled, as, for example, soft-repulsive particles in one dimension, are rather the exception then the rule.

\begin{figure}[t!!!!]{
\includegraphics[width=0.235\textwidth] {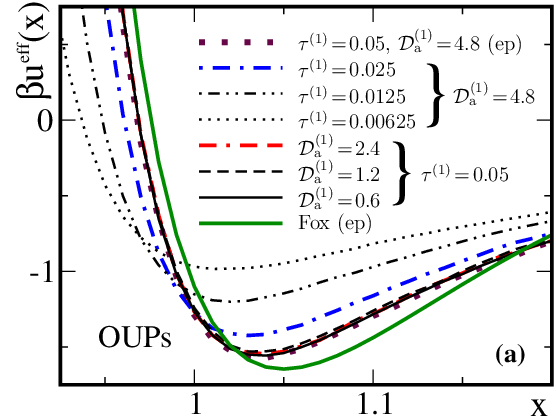} \hfill
\includegraphics[width=0.235\textwidth] {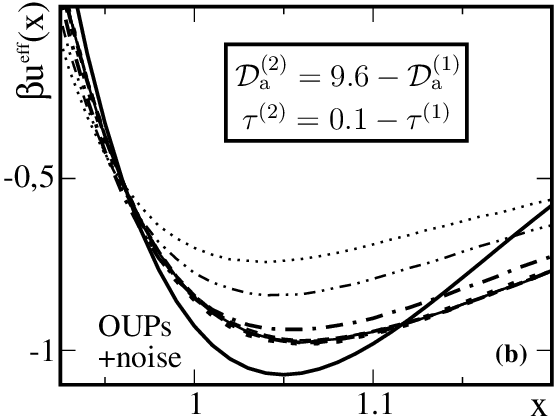}}
\caption{Effective potentials in one dimension from OUPs simulation and the Fox theory, Eq.~\eqref{eq_ueffMIX1d}, between two active particles 
with the parameters $\tau^{(1)}$ and $\Da^{(1)}$ (as labeled) with $\tau^{(2)}\!=\!0.1-\tau^{(1)}$ and $\Da^{(2)}\!=\!9.6-\Da^{(1)}$, 
such that $\bar{\tau}\!=\!0.05$ and $\bar{\mathcal{D}}_\text{a}\!=\!4.8$ (the results are invariant when exchanging the particles, i.e., the labels 1 and 2).
Here we consider only particles with at least one pair of equal parameters ($\tau^{(2)}\!=\!\tau^{(1)}$ or $\Da^{(2)}\!=\!\Da^{(1)}$),
such that the results only depend on $|\Delta\Da|$ or $|\Delta\tau|$, 
respectively. 
The case of all parameters being equal is labeled as (ep).
The Fox theory predicts the same result (ep) in all cases.
We consider a system \textbf{(a)} without and \textbf{(b)} with thermal noise; note the different scale on the vertical axes. 
The legends apply to both subfigures and the colored lines reappear in subsequent figures for comparison.
\label{fig_sameDatau}}
\end{figure}

 Regarding mixtures, we should expect that additional difficulties arise if the term $\Delta\Da\,\Delta E_n(x)$ in Eq.~\eqref{eq_EVgen},
 which is not present in the result based on the UCNA, is positive.
 The  nature of this term can be understood by the following example.
In the present model we can define the passive (Brownian) particle (label 2) in two ways~\cite{activePair}:
   we always require that $E^{(2)}_2\!=\!1$, i.e., $\tau^{(2)}\!=\!0$
 (likewise, in the one-component UCNA, the bare potential of a passive particle can only be recovered when the persistence time is set to zero~\cite{activePair}).
  Obviously, the persistence time $\tau^{(1)}$
   of the active species has to remain finite.
   Then we can set either
 $\Da^{(2)}\!=\!1$ with $I_\text{t}^{(2)}\!=\!0$  or $\Da^{(2)}\!=\!0$ with $I_\text{t}^{(2)}\!=\!1$.
Both definitions result in the same $\mathcal{E}_2$ for an active-passive mixture with arbitrary $\tau^{(1)}$ and $\Da^{(1)}$  only if the term $\Delta\Da\,\Delta E_2(x)$ is present.

\begin{figure}[t!!!!]{
\includegraphics[width=0.235\textwidth] {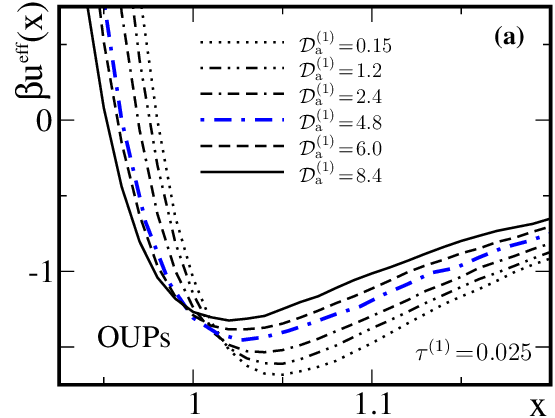} \hfill
\includegraphics[width=0.235\textwidth] {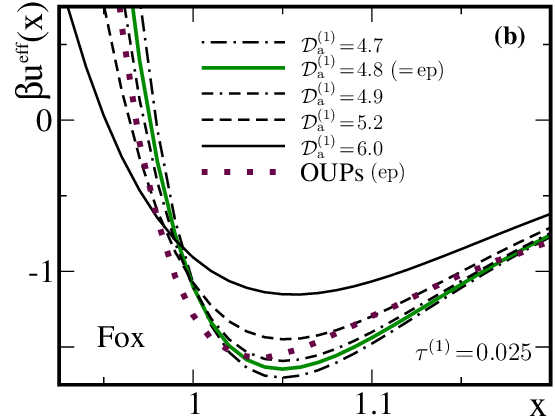} }
\caption{ 
As Fig.~\ref{fig_sameDatau}, but for two particles with fixed $\tau^{(1)}\!=\!0.025$ and thus $\tau^{(2)}\!=\!0.075$.
We qualitatively compare \textbf{(a)} simulations to \textbf{(b)} theory,
where $\Delta\Da$ is chosen much smaller for the theoretical curves,
since the changes for different $\Da^{(1)}$ are more significant and
the curves start to diverge for some $\Da^{(1)}\!<\!4.8$.
 This is a direct consequence of the form of Eq.~\eqref{eq_ueffMIX1d}.
  The colored lines for $\Da^{(1)}\!=\!4.8$ and for all parameters being equal (ep) are the same as in Fig.~\ref{fig_sameDatau}a.
\label{fig_sameDatau2}}
\end{figure}

 Solving Eq.~\eqref{eq_ueffij} we obtain the most general form 
 \begin{align}
\beta u^\text{eff}_{12}(x)=&
\int^x_{-\infty}\!\upd s\,\mathcal{E}_2(s)\,\partial_s\beta u(s)-\ln\!\bigg(\bar{\mathcal{D}}_\text{a}\,\left|\mathcal{E}_2(x)\right|\bigg)
\label{eq_ueffMIX1d}
\end{align}
 of the effective potential for a symmetric bare potential with $\lim_{x\rightarrow\infty}u(x)\!=\!0$.
 For two different species, it is not possible to carry out the integral in general, even if $I_\text{t}^{(1)}\!=\!I_\text{t}^{(2)}\!=\!0$.
 If, in addition, both species either have the same active diffusivity $\Da^{(1)}\!=\!\Da^{(2)}$
 or persistence time $\tau^{(1)}\!=\!\tau^{(2)}$, one recovers an effective potential similar to the one-component result in Eq.~\eqref{eq_ueffUCNA0},
 replacing $\tau^{(1)}\!\rightarrow\!\bar{\tau}$ 
 or $\Da^{(1)}\!\rightarrow\!\bar{\mathcal{D}}_\text{a}$, respectively,
 with the appropriate average parameter.

 Although we restrict ourselves to the one-dimensional case here,
 we briefly extend the above discussion to higher spatial dimensions.
Whereas, the effective potential for a single species is highly accurate~\cite{maggi2015sr,activePair} in one dimension,
implementing the general, higher dimensional, results in would most likely come along with the following caveats (as detailed in appendix~\ref{appendix:cartesian}):
(i) the exact effective potentials can be written in a form similar to Eq.~\eqref{eq_ueffMIX1d}, i.e., the argument of the logarithm follows from the determinant of the effective diffusion tensor~\cite{activePair}, only if 
there is no thermal noise and we assume $\Da^{(1)}\!=\!\Da^{(2)}$ or $\tau^{(1)}\!=\!\tau^{(2)}$;
(ii) we have no rigorous proof that the underlying assumption $\bvec{J}_1\!=\!\bvec{J}_2$ of equal probability currents holds for $\di\!>\!1$;
(iii) already for two particles of the same species, empirical corrections of the effective potential are required
and the quantitative agreement with computer simulation becomes worse with increasing dimensionality.
  However, we stress that most relevant cases (to be discussed later) are consistent with the assumptions under point (i).
Moreover, for a single component, it has been shown~\cite{activePair} that deviations due point (i) are not severe and, despite point (iii), qualitatively correct behavior can be retained.

\subsection{Model calculations \label{sec_ueffresults}}

To test the generalized theory we consider a soft-repulsive bare potential $u(x)\!=\!x^{-12}$ between two particles
and perform computer simulations  of active OUPs~\cite{activePair}, evolving according to Eq.~\eqref{eq_FOXapp1}, as a benchmark for the effective potentials predicted from Fox's approach.
 Since we focus on the one-dimensional case, we can also make quantitative statements
about whether the accuracy of the single-component theory~\cite{marconi2016mp,activePair,maggi2015sr} is maintained if the difference in activity increases.
We checked that the qualitative behavior is not altered when thermal noise is present
and a quantitative comparison to the simulation data is thus analogous to
the single-component case~\cite{activePair}.
An exemplary direct comparison of these two systems is made in Fig.~\ref{fig_sameDatau}.
For the following discussions, we assume $I_\text{t}^{(1)}\!=\!I_\text{t}^{(2)}\!=\!0$
and recall the definitions and sign of relative parameters from Eq.~\eqref{eq_avpars}.

\begin{figure}[t!!!!!]{
\includegraphics[width=0.235\textwidth] {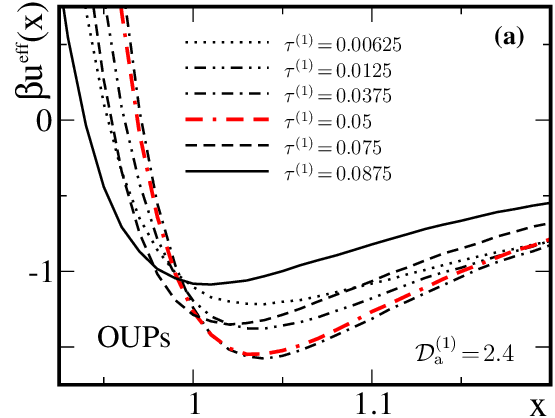} 
 \hfill
\includegraphics[width=0.235\textwidth] {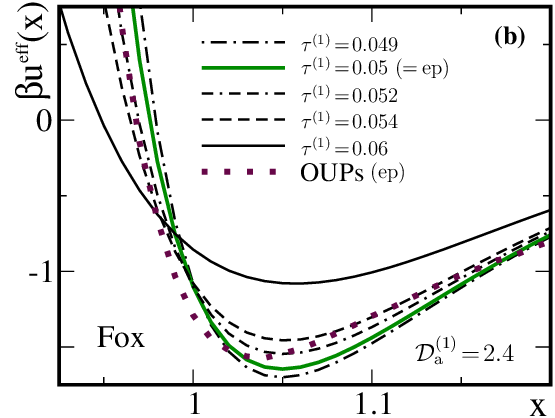}}
\caption{
As Fig.~\ref{fig_sameDatau}, but for two particles with fixed $\Da^{(1)}\!=\!2.4$ and thus $\Da^{(2)}\!=\!7.2$.
As in Fig.~\ref{fig_sameDatau2}, we compare \textbf{(a)} simulations to \textbf{(b)} theory, where the curves diverge for some $\tau^{(1)}\!<\!0.05$.
 The colored lines for $\tau^{(1)}\!=\!0.05$ and for all parameters being equal (ep) are the same as in Fig.~\ref{fig_sameDatau}a.
\label{fig_sameDatau3}}
\end{figure}

The most important theoretical statement of Sec.~\ref{sec_ueffexplicit} is that either for equal $\Da^{(1)}\!=\!\Da^{(2)}\!=\!\bar{\mathcal{D}}_\text{a}$
 or $\tau^{(1)}\!=\!\tau^{(2)}\!=\!\bar{\tau}$ the effective potential of the mixture is equal to that of two identical particles with averaged activity parameters.
According to Fig.~\ref{fig_sameDatau} this prediction is indeed confirmed numerically for the latter case, 
whereas, for equal diffusivities, the effective potential should rather become less attractive with increasing difference $|\Delta\tau|$.
 We will return to this point at the end of this section.

If the two species differ in both activity parameters, the theoretical results do no longer depend only on the average values
due to the term $\Delta\Da\,\Delta E_2(x)$ in Eq.~\eqref{eq_ueffMIX1d}.
For small differences $\Delta\Da$ and $\Delta\tau$, 
we observe in Figs.~\ref{fig_sameDatau2} and~\ref{fig_sameDatau3} that
both theory and simulations predict a deeper minimum of the effective potential when $\Delta\Da\Delta\tau$ is increased.
Further increasing the absolute value of either difference, the theory becomes quantitatively inaccurate.
For $\Delta\Da\Delta\tau<0$ the theoretical curves suggest a rapid decline in the effective attraction,
while much larger differences between the parameters are required to noticeably shift the numerical curves.
For $\Delta\Da\Delta\tau>0$ the theory starts to predict diverging effective potentials, which is qualitatively wrong.

 Interestingly, the behavior of the simulation results in the regime $\Delta\Da\Delta\tau>0$ distinctly depends on the parameters that are changed.
 Fixing $\Delta\tau<0$ and decreasing $\Delta\Da<0$ the curves in Fig.~\ref{fig_sameDatau2}a begin to saturate 
 and the deepest minimum is reached for the minimal $\Delta\Da$  i.e., $\Da^{(1)}\!=\!0$.
 On the other hand, the curve with the deepest minimum in Fig.~\ref{fig_sameDatau3}a at constant $\Delta\Da<0$
 is found for an intermediate $\Delta\tau<0$, 
 whereas for even smaller $\Delta\tau$ the trend inverts. 
 This means that for large absolute differences $|\Delta\tau|$
 in the persistence time, i.e., one species becoming more and more passive, 
 there is no significant attraction between two active OUPs (see also Fig.~\ref{fig_sameDatau}). 

\begin{figure}[t]{

\includegraphics[width=0.2293\textwidth] {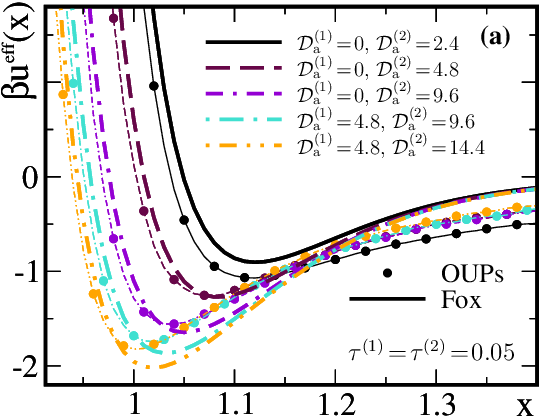} \hfill
\includegraphics[width=0.2293\textwidth] {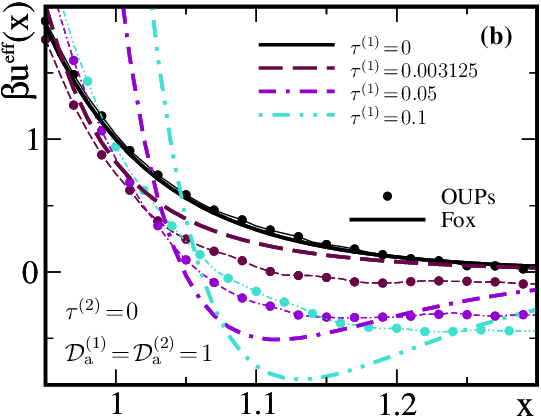}}

\caption{Comparison between the effective potentials from theory (thick lines) and simulations (dots and thin lines) for \textbf{(a)} a fixed persistence time $\tau^{(1)}\!=\!\tau^{(2)}\!=\!0.05$
but different active diffusivities (as labeled)
and \textbf{(b)} an active-passive mixture with fixed $\Da^{(1)}\!=\!\Da^{(2)}\!=\!1$ and the passive $\tau^{(2)}=0$ but different active persistence times (as labeled). 
In this case, the equilibration appears to proceed very slowly and the numerical values only gradually approach 0 for separations larger than shown here.
However, even for large values of $\tau^{(1)}$ there is no attractive well,
suggesting that the fully equilibrated data should reflect the behavior of two passive particles.
\label{fig_tauCOMPARE}}
\end{figure}

The special case of a common persistence time $\tau\!=\!\tau^{(1)}\!=\!\tau^{(2)}$ of both particles is particularly relevant,
since $\tau$ represents the rotational diffusion of all species 
in a mixture of ABPs with different self-propulsion velocities~\cite{faragebrader2015}. 
As noted before, this choice of parameters also yields particularly simple effective potentials,
which are equivalent to the single-component results with the averaged diffusivity $\bar{\mathcal{D}}_\text{a}$ defined according to Eq.~\eqref{eq_avpars}.
Recall from  Fig.~\ref{fig_sameDatau} that also computer simulation results are well represented by those with $\bar{\mathcal{D}}_\text{a}$.
Figure~\ref{fig_tauCOMPARE}a nicely confirms for different activity parameters our expectation
 that for the Brownian mixture under consideration the effective potentials are as accurate as
 those between identical particles in one spatial dimension~\cite{maggi2015sr,activePair}.
 Accordingly, the deviations from the simulation results are most significant for large separations $x$ 
 and increase with increasing average activity $\bar{\mathcal{D}}_\text{a}$.

Another special case, which recently has attracted much interest, is a mixture of an active and a passive Brownian species~\cite{wysocki2016propagatingAP,stenhammar2015activity,vandermeer2017}.
Given the prior results, the
only way to set up a meaningful theoretical description of such a system
is to fix $\Da^{(1)}\!=\!\Da^{(2)}\!=\!1$,
since the persistence times $\tau^{(2)}\!=\!0$ of the passive and $\tau^{(1)}$ of the active species
are different by necessity in the OUPs model.
Choosing the latter as the free activity parameter, as shown in Fig.~\ref{fig_tauCOMPARE}b, makes it difficult to connect to ABPs, 
where the activity should rather be tuned by varying the active diffusivity $\Da^{(1)}$
(depending on the self-propulsion velocity)
than the (non-Brownian) reorientation time $\tau^{(1)}$.

In general, our OUPs simulations indicate that there is no significant effective attraction whenever one of the two particles is passive.
 Putting aside the difficulties with the equilibration, the numerical curves in Fig.~\ref{fig_tauCOMPARE}b
are practically independent of the activity of the second particle.
 Inspired by Percus' test particle approach~\cite{Percus62}, we conjecture  
that the pair distribution in a two-body system including a passive Brownian particle
always reflects the behavior of the bare interaction potential,
regardless of the type and magnitude of self propulsion of the other species. 
This conclusion is also consistent with the behavior of the different curves shown in Figs. 1 and 3a upon increasing $|\Delta\tau|$.
As for these sets of parameters, the theoretical effective potentials in Fig.~\ref{fig_tauCOMPARE}b overestimate the effective attraction.
However, in this special case with one passive particle the simulation result obtained on the two-body level 
obviously does not reflect the behavior of the many-body system.

\section{Conclusions}
\label{Conclusions}

In this paper, we derived the multicomponent generalization of the multidimensional Fox approximation
and applied it to pairs of active particles with different persistence times and diffusivities.
 We argued that the present approach better describes the non-equilibrium behavior of such systems, compared to the UCNA. 
Our analytic results for two particles in one dimension were compared to an exact solution for harmonic potentials
and computer simulations for soft-repulsive interactions,
which both suggest that our theory is satisfactory at linear order in the persistence times.
Explicitly, the formula in Eq.~\eqref{eq_ueffMIX1d} with Eq.~\eqref{eq_EVgen} for the effective interaction potential between two different particles
resemble those for two identical particles with average activity parameters plus an additional term depending on their differences.
This term is important to make correct qualitative predictions close to equilibrium (or for small differences between the parameters)
 but also can be identified as the reason for the wrong or even unphysical predictions beyond the low-activity limit.
 It might be interesting in future work to explore the possibility for an empirical modification of the effective potential
in the spirit of the inverse-$\tau$ approximation introduced in Ref.~\cite{activePair}.

To understand the value of the theory for a many-body system, 
one must think of possible approximations for the effective Fokker-Planck equation~\eqref{eq_FOXappFP11}.
A convenient method to achieve this goal is to construct an effective many-body force from the derived effective potentials~\cite{activePressure}.
Likewise these can be readily implemented in a density functional theory
in order to make explicit predictions of the structure and phase behavior in the steady state~\cite{wittmannbrader2016}.
Keeping this in mind, we also stress that the limitations discussed in Sec.~\ref{sec_ueffresults}
 if the two particles belong to different species,
will eventually only play a minor role on the many-particle level.
Here, the effective pair potential between members of the same species will be equally or, most likely, even more relevant. 
The same is true if one is interested in a mixture in the presence of external forces,
which are also single-body quantities.
It might thus also constitute a fair approximation to drop the term $\Delta\Da\,\Delta E_2(x)$ in Eq.~\eqref{eq_EVgen},
i.e., to simply use the average parameters in any case.

The most obvious application of our effective potentials is also among those of most recent interest.
Following the intuition from passive mixtures, 
the tendency of a two-component system to demix arises from differences in the interactions between the members of each species.
In our case the increased effective attraction within the (more) active species
can be interpreted as the driving force of the demixing process when its activity is increased.
It will thus come to no surprise that an explicit (passive) calculation
will predict that the effective equilibrium state of a mixture of two different active species is demixed/phase separated.
Of course the problem of demixing is ill-defined in one dimension, but the results discussed here will qualitatively be the same in higher dimensions
(if an appropriate correction is employed to avoid possible divergences~\cite{activePair}).

Considering the problem of active depletion it is not as simple to draw conclusion solely from discussing the effective potentials.
The expected attraction between the passive particles does not result directly from the (still passive) effective potential between members of this species.
It is still quite likely that the enhancement of the depletion interaction is also captured in our theory
and can be implicitly accounted for when studying the full behavior of an effective mixture between an active ideal gas (higher effective temperature) and passive colloids.
To show this explicitly, another calculation in the spirit of the Asakura-Oosawa model~\cite{AO} would be necessary.
 Likewise, all other combinations of different passive potentials in the active mixture can be modeled within our approach,
where the effective interactions between members of different species, as studied here, are generally important.

In a nutshell, the presented theoretical framework provides the basis to study active mixtures using methods familiar from equilibrium liquid-state theory,
following the examples elaborated for a single species in arbitrary dimensions. 
Apart from the possibilities discussed above by taking advantage of the effective potentials, 
further work could also address the pressure and interfacial tension~\cite{marconi2016,activePressure},
and an extension of the theory to study dynamical problems~\cite{activePair}.

\section*{Acknowledgements}

R.\ Wittmann and J.\ M.\ Brader acknowledge funding provided by the Swiss National Science Foundation.

\appendix 
\section{The multicomponent Fox approximation}
\label{appendix:foxapproach}

We derive the Fokker-Planck evolution equation of the probability density distribution  associated with
the stochastic differential equation~\eqref{eq_FOXapp1}
by employing a generalized Fox approximation~\cite{faragebrader2015,sharma2016}.
 We consider here the most general case of $N$ particles in $\di$ spatial dimensions,
which does not increase the level of complexity of the following mathematical steps
compared to the special case $N\!=\!2$ and $\di\!=\!1$ discussed in the main text. 
In the chosen notation, each Cartesian coordinate of each particle is considered as an individual component.

By differentiating with respect to time  the probability density distribution 
\begin{align}
f_N(\{y_\upalpha\},t)=\int D[\{\chi_\upalpha\}]P_N[\{\chi_\upalpha\}]\,\prod_{\upalpha=1}^{\di N}\delta(y_\upalpha-x_\upalpha(t))
\label{eq_distribution}
\end{align}
associated the stochastic processes $x_\upalpha(t)$, defined in Eq.~\eqref{eq_FOXapp1} of the main text,
we obtain the following equation:
\begin{align}
&\frac{\partial f_N(\{y_\upalpha\},t)}{\partial t}\cr&=
-\sum_{\upbeta=1}^{\di N}\frac{\partial}{\partial y_\upbeta} \Biggl(
D_\text{t}\beta F_{\upbeta}(\{y_\upalpha\})f_N(\{y_\upalpha\},t)\cr &\ \ \ \ \ \ +\int D[\{\chi_\upalpha\}]P_N[\{\chi_\upalpha\}] 
\bigg(\prod_{\upalpha=1}^{\di N}\delta(y_\upalpha-x_\upalpha(t))\bigg)\chi_{\upbeta}(t) \Biggr)\, .\cr
\label{eq_FOXappFP0}
\end{align}
The first term in Eq.~\eqref{eq_FOXappFP0}
stems from the deterministic part of the evolution, whereas the second term 
accounts for the noise contribution and  is calculated as follows.

We first use the Novikov theorem and
the explicit form of the noise correlation in Eq.~\eqref{eq_OUPsCORRapp} of the main text
and rewrite the last term in Eq.~\eqref{eq_FOXappFP0} as:
\begin{align}
&\int D[\{\chi_\upalpha\}]P_N[\{\chi_\upalpha\}]\left(\prod_{\upalpha=1}^{\di N}\delta(y_\upalpha-x_\upalpha(t))\right)\chi_\upbeta(t)\cr
&= - \sum_{\upgamma=1}^{\di N}\int\upd t'C_{\upbeta\upbeta}(t-t')\frac{\partial}{\partial y_\upgamma} \int D[\{\chi_\upalpha\}]P_N[\{\chi_\upalpha\}]
\ \cr&\ \ \ \ \ \ \ \ \times\left(\prod_{\upalpha=1}^{\di N}\delta(y_\upalpha-x_\upalpha(t))\right)
\frac{\delta x_\upgamma( t)}{\delta\chi_\upbeta(t')} \, .
\label{eq_FOXapp3}
\end{align}
 In order to evaluate the response function (the last factor) on the right-hand side of the above expression, we use again Eq.~\eqref{eq_FOXapp1} and find
\begin{align}
\frac{\delta\dot{x}_\upgamma(t)}{\delta\chi_\upbeta(t')}=D_\text{t}\sum_{\updelta=1}^{\di N}
\frac{\partial\beta F_\upgamma(\{x_\upalpha(t)\})}{\partial x_\updelta(t)}\frac{\delta x_\updelta(t)}{\delta\chi_\upbeta(t')}+\delta_{\upbeta\upgamma}\delta(t-t')\,.
\label{eq_FOXapp4}
\end{align}
 The formal solution of Eq.~\eqref{eq_FOXapp4} with the initial condition 
\begin{eqnarray}
\left[\frac{ \delta x_\upgamma(t)}{\delta \chi_\upbeta(t')} \right]_{t=t'}=\delta_{\upbeta\upgamma} 
\label{has12}
\end{eqnarray}
 is given (for $t>t'$) by  the tensor
 \begin{align}
\frac{\delta x_\upgamma(t)}{\delta\chi_\upbeta(t')}=&
\left(\exp\int_{t'}^t\upd s\; \boldsymbol{\mathfrak{F'}}(s)\right)_{\upgamma\upbeta}\Theta(t-t')
\cr&
\approx \left(e^{(t-t')\,\boldsymbol{\mathfrak{F'}}(t)}\right)_{\upgamma\upbeta} \Theta(t-t')\,.
\label{eq_FOXapp5}
\end{align}
In the second step, we expanded the integral~\cite{faragebrader2015} in the exponent up to linear order in $(t-t')$
and we introduced $\boldsymbol{\mathfrak{F'}}(t)\simeq\boldsymbol{\mathfrak{F'}} [\{x_\upalpha(t)\}]$ with the components 
$\boldsymbol{\mathfrak{F'}}_{\upgamma\upbeta}=D_\text{t}\partial \beta F_\upgamma/\partial x_\upbeta$.

To shorten the notation we indicate the average of a function ${\cal O}(\{x_\upalpha(t)\})$ as
\begin{align}
\langle {\cal O}(\{x_\upalpha(t)\}) \rangle\equiv \int D[\{\chi_\upalpha\}] P_N[\{\chi_\upalpha\}] {\cal O}(\{x_\upalpha(t)\})
\end{align}
 Now we use Eq.~\eqref{eq_FOXapp5} to rewrite Eq.~\eqref{eq_FOXapp3}  as 
\begin{align}
&
-\frac{\partial}{\partial y_\upgamma}
\sum_{\upgamma=1}^{\di N}\int_0^t \upd t'C_{\upbeta\upbeta}(t-t')
\Bigl\langle \prod_{\upalpha=1}^{\di N}\delta(y_\upalpha-x_\upalpha(t))
\frac{\delta x_\upgamma( t)}{\delta\chi_\upbeta(t')} \Bigr\rangle \nonumber \\
&
\approx
-\sum_{\upgamma=1}^{\di N}\frac{\partial}{\partial y_\upgamma} f_N( \{y_\upalpha\},t) \int_0^t \upd t'C_{\upbeta\upbeta}(t-t')
 \nonumber\\&\ \ \ \ 
\times \Bigl\langle  \left(e^{(t-t')\,\boldsymbol{\mathfrak{F'}}(t)}\right)_{\upgamma\upbeta}  \Bigr\rangle  \,,
\label{expression}
\end{align}
where, according to Eq.~\eqref{eq_distribution}, $f_N \equiv \big\langle \prod_{\upalpha}\delta(y_\upalpha-x_\upalpha(t)) \big\rangle$
and we approximated the average of the product in the in the first line by the 
product of the averages.

Using the explicit correlator~\eqref{eq_OUPsCORRapp}
and further approximating the average of the exponential 
 with the exponential of the average the  integral featuring in Eq. ~\eqref{expression} becomes:
 \begin{align}
 && \int_0^t \upd t'  \frac{D_\text{a}^{(\upbeta)}}{\taua^{(\upbeta)}}e^{-\frac{|t\!-\!t'|}{\taua^{(\upbeta)}}}  
\times  \left(e^{(t-t')\, \Bigl\langle \boldsymbol{\mathfrak{F'}}(t)   \Bigr\rangle}  \right)_{\upgamma\upbeta}   \,
\nonumber\\&&
\approx 
D_\text{a}^{(\upbeta)} \Bigl(
{\bf I} -\taua^{(\upbeta)} \Bigl\langle \boldsymbol{\mathfrak{F'}}(t)   \Bigr\rangle 
 \Bigr)^{-1}_{\upgamma\upbeta}  \,,
\label{eq_FOXapp6}
\end{align}
where we took the small $\taua^{(\upbeta)}$ limit in the integral.
Putting together
\begin{align}
&\frac{\partial f_N(\{y_\upalpha\},t)}{\partial t}\cr&=
-\sum_{\upbeta=1}^{\di N}\frac{\partial}{\partial y_\upbeta} \Biggl(
D_\text{t}\beta F_{\upbeta}(\{y_\upalpha\})f_N(\{y_\upalpha\},t)\cr &\ \ \ \ \ \ 
-D_\text{a}^{(\upbeta)} \sum_{\upgamma=1}^{\di N}  \frac{\partial}{\partial y_\upgamma} f_N( \{y_\upalpha\},t) \Bigl(
{\bf I} -\taua^{(\upbeta)} \Bigl\langle \boldsymbol{\mathfrak{F'}}(t)   \Bigr\rangle 
 \Bigr)^{-1}_{\upgamma\upbeta} \Biggr). \cr
\label{eq_FOXappFP10}
\end{align}
Such a formula can be recast under the form of Eq.~\eqref{eq_FOXappFP11} given in the main text,
where, in order to reduce the notational burden, we identify the symbol of the coordinates $y_\upalpha$ with that of the stochastic processes $x_\upalpha(t)$,
which are formally equivalent, according to Eq.~\eqref{eq_distribution}.

The Fokker-Planck equation, Eq.~\eqref{eq_FOXappFP10}, can be written in the compact form
\begin{align}
\frac{\partial }{\partial t} f_N(\{y_\upalpha\},t)={\cal L}_\text{FP} f_N(\{y_\upalpha\},t)\,,
\end{align} 
where ${\cal L}_\text{FP}$ represents the Fokker-Planck operator. 
It follows that the average of an observable ${\cal O}(\{y_\upalpha\})$ evolves according to
\begin{align}
\frac{\partial }{\partial t}\langle {\cal O}(\{y_\upalpha\},t)\rangle =\langle{\cal L}^{\dagger}_\text{FP} {\cal O}(\{y_\upalpha\},t)\rangle\,,
\end{align} 
where ${\cal L}^{\dagger}_\text{FP} $ is the adjoint operator of the Fokker-Planck operator ${\cal L}_\text{FP} $ and
$\langle {\cal O}(\{y_\upalpha\},t)\rangle \equiv\int \upd \{y_\upalpha\} f_N(\{y_\upalpha\},t) {\cal O}(\{y_\upalpha\})$.

\section{Theory for $N$ particles in $\di$ dimensions \label{appendix:cartesian}}

For completeness, we restate in this appendix the most important results of the main text in a more general fashion valid in higher spatial dimensions $\di$.
For this purpose, and to make connection to the notation employed in Refs.~\cite{activePair,activePressure},
it is convenient to rewrite Eq.~\eqref{eq_FOXappFP11} of the main text, using Cartesian coordinates and component-wise notation, as.
\begin{align}\!\!\!\!
\frac{\partial f_N(\rr^N,t)}{\partial t}&=
- D_\text{t}\sum_{\upsilon=1}^\kappa\sum_{k_\upsilon}^{N_\upsilon} \nab_{k_\upsilon}\cdot \bvec{J}_{k_\upsilon}\,,
\label{eq_FOXFP}\\
\bvec{J}_{k_\upsilon}(\rr^N,t)&=\beta\bvec{F}_{k_\upsilon} f_N-\sum_{\mu=1}^\kappa\sum_{i_\mu=1}^{N_\mu}\nab_{i_\mu}\cdot(\mathcal{D}_{i_\mu k_\upsilon}f_N)\,,
\label{eq_ss1app}\\
\mathcal{D}_{{i_\mu}{k_\upsilon}}(\rr^N) &=
\boldsymbol{1}\delta_{{i_\mu}{k_\upsilon}} I_\text{t}^{({\upsilon})}+
\Da^{({\upsilon})}\,\Gamma_{{i_\mu}{k_\upsilon}}^{-1}(\rr^N)\,,
\label{eq_Deffapp}\\
\Gamma_{{i_\mu}{k_\upsilon}}(\rr^N)&=\boldsymbol{1}\delta_{{i_\mu}{k_\upsilon}}+\tau^{(\upsilon)}\nab_{k_\upsilon}\nab_{i_\mu}\!
\sum_{\nu=1}^\kappa\smash{\left.\sum_{{j_\nu}=1}^{N_\nu}\right.}^{\!\mathlarger{\mathlarger{'}}} \beta u(\rr_{i_\mu},\rr_{j_\nu})\,,
\label{eq_Gamma0app}
\end{align}
where we first sum over the different particle species $\mu$, so that $N=\sum_\mu N_\mu$.
Then we sum over particles $i_\mu$ of each species, which all have the same persistence times $\tau^{(\mu)}$, active diffusivities $\Da^{(\mu)}$
and characteristic functions $I_\text{t}^{(\mu)}$ of thermal noise. 
 The primed sum excludes the particle in the first argument of the following function.

Equations \eqref{eq_FOXFP}-\eqref{eq_Gamma0app} represent the full extension of Fox's approximation~\cite{fox1,fox2} for one-component active
fluids~\cite{faragebrader2015,sharma2016,activePair} to the multicomponent case where the fluid contains different species and each species is subjected to a different active (colored) noise. 
In general, the effective diffusion tensor $\mathcal{D}_{{k_\upsilon}{i_\mu}}$ is not symmetric, i.e., 
$\mathcal{D}_{{k_\upsilon}{i_\mu}}\neq\mathcal{D}_{{i_\mu}{k_\upsilon}}$, if the particles belong to two different species. 
Regarding the structure of Eqs.~\eqref{eq_FOXFP} and~\eqref{eq_ss1app}, transposing $\mathcal{D}_{{k_\upsilon}{i_\mu}}$ changes the components of the probability current but not the overall time evolution of the probability distribution.
With Eq.~\eqref{eq_Deffapp} coupling the $N$ particle positions in a non-trivial way, 
the general theory does not provide any useful simplification of Eq.~\eqref{eq_FOXapp1} at this stage.

One possibility to arrive at a computationally tractable many-particle theory is to naively define an effective current
\begin{equation}
\tilde{\bvec{J}}_{k_\upsilon}(\rr^N,t)= \sum_{\nu=1}^\kappa\smash{\left.\sum_{{j_\nu}=1}^{N_\nu}\right.}^{\!\mathlarger{\mathlarger{'}}}\mathcal{D}_{i_\mu k_\upsilon}
\cdot\left(\beta\tilde{\bvec{F}}^\text{eff}_{i_\mu} f_N-\nab_{i_\mu} f_N\right)
\label{eq_jeff}
\end{equation}
with the (approximate) pairwise additive effective force
\begin{equation}
 \tilde{\bvec{F}}^\text{eff}_{i_\mu}=-\nab_{i_\mu}\sum_{\nu=1}^\kappa\sum_{j_\nu=1}^{N_\nu}u^\text{eff}_{\mu\nu}(\rr_{i_\mu},\rr_{j_\nu})
\end{equation}
constructed from the effective potentials $u^\text{eff}_{\mu\nu}(r)$, as defined for $\di\!=\!1$ in Sec.~\ref{sec_ueffexplicit} of the main text.
Since there is no clean definition of an exact effective many-body force~\cite{faragebrader2015,activePair} in the present case, 
the form of Eq.~\eqref{eq_jeff} has been adopted from that for a single-component system~\cite{activePressure}.
For a mixture of $N=2$ particles, it has been explicitly discussed in the main text, that such a separation in not exact.
In the words of Ref.~\cite{activePressure}, it is not possible to write the steady-state condition in a thermodynamical version~\cite{activePressure}.
However, the definition in Eq.~\eqref{eq_jeff} comes along the approximation that there exists a current-free steady state 
with the probability distribution $f_N\propto\exp(-\frac{1}{2}\sum_{\mu\nu}\sum_{i_\mu j_\nu}\beta u^\text{eff}_{\mu\nu})$ and 
therefore closed theories can be constructed borrowing methods from equilibrium.

To derive the effective pair potentials $u^\text{eff}_{\mu\nu}$ in dimensions higher than one, we
restate the two-body steady-state condition, Eq.~\eqref{eq_ss1ld}, in Cartesian coordinates
\begin{equation}
\nab\cdot\bvec{J}_1-\nab\cdot\bvec{J}_2=0\,. 
\label{eq_ss1ldAPP}
\end{equation}
Explicitly, the two current vectors are given by the multidimensional version of Eq.~\eqref{eq_j2},
which can be easily generalized by replacing the scalar components of the one-dimensional effective diffusion tensor with Eq.~\eqref{eq_Deffapp}.
From now on we focus on the solutions of $\bvec{J}_1=\bvec{J}_2+\bvec{X}$, the integral of Eq.~\eqref{eq_ss1ldAPP},
and conjecture that the vector field $\bvec{X}$ vanishes.
This is shown explicitly in Sec.~\ref{sec_twopartcurrent} for a one-dimensional system.

For $\di\!>\!1$, the $2\di\times2\di$ friction matrix (of two particles of the same species) from Eq.~\eqref{eq_Gamma0app} has two distinct Eigenvalues
\begin{align}
 E^{(\mu)}_n(r)=1+2\tau^{(\mu)} r^{n-2}\partial^n_r\beta u(r)
 \label{eq_EVapp}
\end{align}
with $n\in\{1,2\}$, from which, following Eq.~\eqref{eq_EVgen} of the main text, we can define the general Eigenvalues $\mathcal{E}_n$ 
of $\mathcal{D}_\text{m}:=\frac{1}{2}\big(\mathcal{D}_{11}-\mathcal{D}_{21}+\mathcal{D}_{22}-\mathcal{D}_{12}\big)$, obtained from the
effective diffusion tensor from Eq.~\eqref{eq_Deffapp}.
As a general result, we propose to write the effective potentials in the form
 \begin{align}
\beta u^\text{eff}_{12}(r)=&
\int^r_{\infty}\!\upd s\,\mathcal{E}_2(s)\,\partial_s \beta u(s)-\ln\!\left(\bar{\mathcal{D}}^{\,3}_\text{a}\,\left|\mathcal{E}_2(r)\right|\left|\mathcal{E}_1(r)\right|^{\di-1}\right),
\label{eq_ueffMIX1dAPP}
\end{align}
where the closed expression of the second term
stems from the approximation 
$\mathcal{D}^{-1}_\text{m}\cdot\nab\cdot\mathcal{D}_\text{m}\approx\nab\ln|\det\mathcal{D}_\text{m}|$.
This step is always exact in one dimension, compare the conversion from Eq.~\eqref{eq_psi2} to Eq.~\eqref{eq_ueffMIX1d} of the main text, and
 we expect that it results only in minor deviations in higher dimensions, based on the similar conclusion drawn for a single component~\cite{activePair}. 

In the absence, $I_\text{t}^{(1)}\!=\!I_\text{t}^{(2)}\!=\!0$, of thermal noise
and for one parameter $\Da^{(1)}\!=\!\Da^{(2)}$ or $\tau^{(1)}\!=\!\tau^{(2)}$
being equal in both species, the first term in Eq.~\eqref{eq_ueffMIX1dAPP} can be integrated
and the second term is no longer approximate.
We then find for these special cases
\begin{align}
\beta u^\text{eff}_{12}(r)&=\frac{\upbeta u(r)+\bar{\tau}\left(\partial_r\beta u(r)\right)^2}{\bar{\mathcal{D}}_\text{a}}-\ln\!\left(\left|\bar{E}_2(r)\right|\left|\bar{E}_1(r)\right|^{\di-1}\right) \ \ 
\label{eq_ueffUCNA0app}
\end{align}
with the average parameters defined in Eq.~\eqref{eq_avpars}, as in one dimension.

\end{document}